\documentclass[epj,draft,referee]{svjour}

\usepackage{graphics,pstricks}

\begin{document}

\title{Direct photon emission from hadronic sources:\\Hydrodynamics vs.\ %
Transport theory}
\author{Bj\o{}rn B\"auchle \inst{1}\inst{2} \and Marcus Bleicher\inst{2}
 }

\institute{Frankfurt Institute for Advanced Studies, Ruth-Moufang-Str.\ 1,
60438 Frankfurt am Main \and Institut f\"ur Theoretische Physik, Johann
Wolfgang Goethe-Universit\"at, Max-von-Laue-Str.\ 1, 60438 Frankfurt am
Main, Germany}

\date{Received: date / Revised version: date}

\abstract{
Direct photon emission in heavy-ion collisions is calculated within the
relativistic microscopic transport model UrQMD. We compare the results from
the pure transport calculation to a hybrid-model calculation, where the
high-density part of the evolution is replaced by an ideal 3-dimensional
fluiddynamic calculation. The effects of viscosity, present in the transport
model but neglected in ideal fluid-dynamics, are examined. We study the
contribution of different production channels and non-thermal collisions to
the spectrum of direct photons. Detailed comparison to the measurements by
the WA~98-collaboration are undertaken.
}

\maketitle

\section{Introduction}\label{sec:intro}

The creation and study of high-density and -temperature nuclear matter is
the major goal of heavy-ion experiments. If the energy density in the
reaction is high enough, a state of quasi-free partonic degrees of freedom,
a Quark-Gluon-Plasma (QGP) \cite{Harris:1996zx,Bass:1998vz}, may be formed.
Recent experimental observations at the Relativistic Heavy Ion Collider
(BNL-RHIC) like e.g.\ strong jet quenching and high elliptic flow hint to
the creation of a strongly coupled QGP (sQGP)
\cite{Adams:2005dq,Back:2004je,Arsene:2004fa,Adcox:2004mh}. Possible
evidence for the creation of this new state of matter has also been put
forward by collaborations at the Super Proton Synchrotron (CERN-SPS), as for
instance the step in the mean transverse mass excitation function of
protons, kaons and pions and the enhanced $K^+/\pi^+$-ratio \cite{:2007fe}.

Unfortunately, it is not yet possible to describe the time evolution of the
produced matter from first principle Quantum Chromodynamics (QCD). Neither
can the hot part of a collision be observed directly. Therefore
well-developed dynamical models to describe the space-time evolution of
nuclear interactions are needed. Among those approaches is relativistic
transport theory
\cite{Bass:1998ca,Bleicher:1999xi,Molnar:2004yh,Xu:2004mz,Lin:2004en,Burau:2004ev}.
In this kind of microscopic description, the hadronic and/or the partonic
stage of the collision can be described under certain approximations. E.g.\
most transport models cannot describe collisions with more than two incoming
particles restricting the applicability to low particle densities, where
multi-particle interactions are less important. Some attemps to include
multi-particle interactions do exist
\cite{Barz:2000zz,Xu:2004mz,Larionov:2007hy,Bleibel:2006xx,Bleibel:2007se}, but this
field of study is still rather new. The coupling of a partonic phase with a
hadronic phase imposes one more problem on transport models. The microscopic
details of that transition are not known. There are several attempts to
address the problem of microscopic hadronization
\cite{Lin:2004en,Andersson:1977xs,Ellis:1995pe,Traxler:1998bk,Biro:1998dm,Hofmann:1999jx}.
One caveat in transport approach is that all microscopic scatterings are
explicitly in the model and therefore the cross-sections for all processes
must be known or extrapolated. However, for most processes no experimental
data are available, and a large fraction of cross-sections have to be
calculated or parametrized by additional models.

Relativistic, non-viscous, perfect fluid-dynamics is a different approach to
explore the space-time evolution of a heavy-ion collision. It constitutes a
macroscopic description of the matter that is created, assuming that at
every time and in every place it is in perfect local thermal equilibrium.
This assumption can only be true if the matter is sufficiently dense, so in
the late stages of a heavy-ion collision fluid-dynamics looses
applicability. An advantage is that in the dense stages, hydrodynamics can propagate any
kind of matter, and also allows for transitions between two types of matter,
e.g.\ QGP and hadron gas, if an appropiate Equation of State (EoS) is
provided. Fluid-dynamics can therefore be used to study hadronic and
partonic matter in one common model.

The restrictions of this kind of model can be loosened as well. By
introducing viscosity and heat conductivity, perfect thermal equilibrium
does not have to be present at any point. However, even with second order
corrections the matter has to be close to equilibrium.

Input to solve the hydrodynamic differential equations are the boundaries,
i.e.\ the initial state (the distributions of all relevant densities and
currents at the time the evolution starts), the Equation of State providing
the pressure as function of the energy and baryon densities, that describes
the behaviour of the matter that is considered, and the freeze-out
hypersurface.

\bigskip

Out of the many possible observables, electromagnetic probes have the
advantage of leaving undisturbed: once they are created, they will escape
the reaction zone, due to negligible rescattering cross-sections.
Besides dileptons, direct photon emission is therefore of greatest interest
to gain insight into the early, hot and therefore possibly partonic stages
of a reaction.  Direct photons are distinguished from the mass of photons as
those coming from collisions and not decays.

Unfortunately, most photons in heavy-ion collisions come from hadronic
decays in the very late stages, mostly $\pi^0 \rightarrow \gamma\gamma$.
These decay-photons impose a serious challenge to the experimentalists when
they try to obtain the spectra of direct photons. Up to now, several
experiments have tried to obtain the spectra of direct photons: Helios,
WA~80 and CERES (all at CERN-SPS) could publish upper limits, while WA~98
(CERN-SPS) and PHENIX (BNL-RHIC) have published data points for direct
photons.

Approaches to the theoretical description of direct photon spectra include
calculations with perturbative Quan\-tum-Chromo-Dynamics (pQCD), the
application of thermal rates and microscopic cross-sections, all of which
have their area of applicability.

Calculations based on pQCD describe the photon data in proton-proton
collisions very well and, if scaled by the number of binary collisions, also
those in heavy-ion reactions. The range of applicability is limited to high
transverse momentum $p^{\gamma}_\bot \gg 1~{\rm GeV}$.

Thermal rates can only be applied if the assumption of local thermal
equilibrium is fulfilled. Scattering rates can then be calculated by folding
the particle distribution functions of the participating particle species
with the respective cross-sections.  This framework can be applied to either
static models, simplified hydrodynamics-inspired models such as the blast
wave model and to full fluid-dynamic calculations. The space-time evolution
of a reaction as predicted by microscopic theories can be averaged over in
order to apply thermal rates to the coarse-grained
distributions~\cite{Huovinen:2002im}.

The application of microscopic cross-sections can only be undertaken in a
model where all microscopic collisions are known. That limits the field of
use to transport models. There have been several calculations for photon
spectra from transport models~\cite{Dumitru:1998sd,Bratkovskaya:2008iq}

In this paper we first introduce the dynamic models used for the
calculations. Then, we explain the cross-sections and thermal rates used for
photon emission as well as the mechanisms for doing so. Following that, the
results of our calculations are presented and compared to results from the
WA~98-collaboration. Finally, after summary and conclusions, an outlook to
further planned studies is given.

\section{UrQMD hybrid model}\label{sec:Hybrid}

UrQMD (Ultra-relativistic Quantum Molecular Dynamics) is a microscopic
transport model. It includes all ha\-drons and resonances up to masses $m =
2.2~{\rm GeV}$ and at high energies can excite and fragment strings. The
cross-sections are either calculated via detailed balance or para\-metrized
by the additive quark model (AQM), if no experimental values are available.
At high parton momentum transfers, PYTHIA is employed for pQCD scatterings.
UrQMD has been used by Dumitru {\it et al} to study direct photon
emission earlier~\cite{Dumitru:1998sd}.


In this work, we combine and compare the two models mentioned above to
describe the space-time evolution of a heavy-ion reaction: for the
unequilibrated initial state and the low-density final state, the
microscopic transport model UrQMD is used, whereas the high-density part of
the reaction is modelled using ideal 3+1-dimensional fluid-dynamics. For
details of this model see~\cite{Petersen:2008dd}. During the evolution, two
mappings have to be performed: After the incoming nuclei have passed each
other, the baryon-number-, energy- and momentum-densities are smoothed and
put into the hydro calculation. After the local rest frame energy density
has dropped below a threshold value of $\epsilon_{\rm crit} = 730~{\rm
MeV}/{\rm fm}^3\,(\approx 5 \epsilon_0)$ in every point, particles are
created from the densities by means of the Cooper-Frye formula and
propagation is continued in UrQMD. The effect of changing those parameters
is studied in \cite{Petersen:2008dd} and found to be small.

For the present studies, the EoS used in the hydrodynamic part resembles a
free hadron gas with the same degrees of freedom that the transport part
(i.e., UrQMD) has. Since strings are expected to be created in scatterings
with very high center-of-mass energies, and since hydrodynamics only
describes soft scatterings, they are not included in the EoS.

By this choice, one can study the impact of different underlying dynamics of
the reaction on the spectra of photons. This allows to disentangle which
changes are due to different physical assumptions. In further work, we plan
to explore to the EoS, i.e, changes to the assumptions of the underlying
medium (chirally restored or deconfined matter).

\section{Photon emission from hadronic sources}\label{sec:Photons}

In both models, hybrid and pure transport, photon emission is calculated
perturbatively. That means, the evolution of the underlying event is not
changed by the emission of photons. This is justified as the emission
probability for photons is extremely small.

\subsection{Photons from UrQMD}\label{sec:Photons:UrQMD}

In the transport part of our calculations for each scattering the
cross-section for photon production is calculated. Cross-sections are taken
from Kapusta {\it et al}~\cite{Kapusta:1991qp} and Xiong {\it et
al}~\cite{Xiong:1992ui}. In the former work, the photon self-energy from a
Lagrangian involving the pion, rho and photon-fields has been taken as basis
for the calculations:
\begin{equation} \mathcal{L} = |D_\mu \Phi|^2 - m_\pi^2 |\Phi|^2 -
\frac{1}{4} \rho_{\mu\nu} \rho^{\mu\nu} + \frac{1}{2} m_\rho^2 \rho_\mu
\rho^\mu - \frac{1}{4} F_{\mu\nu} F^{\mu\nu} \label{eq:kapusta_lagrangian}
\end{equation}
(for details the reader is referred to \cite{Kapusta:1991qp}). The coupling
of the $\rho$- to the $\pi$-meson is characterized by the coupling constant
$g_\rho$, which is inferred from the decay rate $\rho \rightarrow \pi\pi$
with $g_\rho^2 / 4\pi = 12 \Gamma \omega_0^3 / (m_\rho p_0^3)$. Here, $p_0$
and $\omega_0$ are the momentum and energy of a decay-$\pi$ in the rest
frame of the $\rho$ and $\Gamma$ is the decay-width.  Kapusta {\it et al}
argue that $\Gamma$ is experimentally measured and therefore effectively
includes certain higher-order effects such as vertex corrections.

From this Lagrangian, they calculate cross-sections for the following
processes:
\begin{eqnarray} \pi^\pm \pi^\mp & \rightarrow & \gamma \rho^0,
\label{eq:xs:picpicrho}\nonumber\\ \pi^\pm \pi^0   & \rightarrow & \gamma
\rho^\pm, \label{eq:xs:picpi0}\nonumber\\ \pi^\pm \rho^0  & \rightarrow &
\gamma \pi^\pm, \label{eq:xs:picrh0}\nonumber\\ \pi^\pm \rho^\mp&
\rightarrow & \gamma \pi^0, \label{eq:xs:picrhc}\nonumber\\ \pi^0 \rho^\pm&
\rightarrow & \gamma \pi^\pm, \label{eq:xs:pi0rhc}\nonumber\\ \pi^\pm
\pi^\mp & \rightarrow & \gamma \gamma.  \label{eq:xs:picpicgam}\nonumber
\end{eqnarray}
The last of these is suppressed by an additional factor of $\alpha$ with
respect to the others and is therefore not expected to contribute to the
photon spectra significantly. Kapusta {\it et al} also quote cross-sections
involving the $\eta$-meson ($\pi^\pm \pi^\mp \rightarrow \gamma \eta$ and
$\pi^\pm \eta \rightarrow \gamma\pi^\pm$), but they are omitted in our
calculations.

From the paper by Xiong {\it et al}, we deduce the cross-sections for the
processes
\begin{eqnarray} \pi \rho \rightarrow a_1 \rightarrow \gamma \pi,
\label{eq:xs:a1}\nonumber \end{eqnarray}
averaged over all possible charge combinations. This channel is not included
in Kapusta {\it et al}.

In order to obtain photon spectra, all scatterings that happen during the
transport phase are examined. For every scattering that can produce photons,
the corresponding fraction of a photon
\begin{equation} N^{\rm t}_\gamma = \frac{\sigma_{\rm em}}{\sigma_{\rm tot}}
\label{eq:Ngamma_trans} \end{equation}
is produced. In this formula, $\sigma_{\rm tot}$ is the sum of the total
hadronic cross-section for a collision with these ingoing particles (known
from UrQMD) and the electromagnetic cross-section $\sigma_{\rm em}$ as
calculated by the formul\ae{} from the abovementioned papers.

This number of photons $N^{\rm t}_\gamma$ is then distributed in the solid
angle by the angular distributions as given by $d\sigma / dt$, so that a
whole distribution of fractional photons per collision instead of only one
(complete) photon every $\sim$ 10,000 collisions is created. This allows us to have
a much better statistics with less hadronic events. The small cross-sections
for the processes we consider justify this perturbative ansatz.

\subsection{Photons from hydrodynamics}\label{sec:Photons:Hydro}

For the hydro-evolution we also produce fractional photons at every cell of
the hydrodynamic calculation. We use the parametrizations by Turbide, Rapp
and Gale~\cite{Turbide:2003si}. To obtain these, they start with an
effective non-linear $\sigma$-model Lagrangian with the vector and axial
vector fields implemented as massive gauge fields of the chiral $U(3)_L
\times U(3)_R$ symmetry. They calculate a different set of rates than
cross-sections calculated by Kapusta {\it et al}, namely:
\begin{eqnarray} \pi \pi & \rightarrow & \gamma \rho,
\label{eq:r:pipirh}\nonumber\\ \pi \rho & \rightarrow & \gamma \pi,
\label{eq:r:pirhpi}\nonumber\\ \pi K^\ast & \rightarrow & \gamma K,
\label{eq:r:pikastka}\nonumber\\ \pi K & \rightarrow & \gamma K^\ast,
\label{eq:r:pikakast}\nonumber\\ \rho K & \rightarrow & \gamma K,
\label{eq:r:rhkaka}\nonumber\\ K^\ast K & \rightarrow & \gamma \pi.
\label{eq:r:kastka}\nonumber \end{eqnarray}
The rate for $\pi \rho \rightarrow \gamma \pi$ contains the process $\pi
\rho \rightarrow a_1 \rightarrow \gamma \pi$. The rates involving $K$ and
$K^\ast$ are included in our calculation albeit they are negligible compared
to the others. Turbide {\it et al} also calculate the emission rate from
$\rho$-decay ($\rho \rightarrow \pi \pi \gamma$), but this is not taken into
account for the present work, because we want to restrict ourselves to
collisional photons.

For every cell, we generate one fractional photon, its fraction being the
integral over the invariant rate:
\begin{equation} N^{\rm h}_\gamma = \Delta V \Delta t \int \frac{d^3 p}{E}\,
E \frac{dR}{d^3p}, \label{eq:Ngamma_hyd} \end{equation}
where $\Delta V$ and $\Delta t$ are cell volume and time step, respectively,
and $E dR / d^3p$ is the rate for photon emission as given by Turbide {\it
et al}. 

To get photon spectra with the right lab-frame distribution out of the
(spherically symmetric) local-rest-frame distribution, a
lorentz-transformation has to be applied to the generated photons.

We choose the invariant generalization of the energy $p_\mu u^\mu$ according
to the distribution functions implied by the rates. Then we choose a $p^\mu$
which, together with the cell's flow velocity $u^\mu$, results in the
desired $p_\mu u^\mu$.

One fractional photon is then created in the direction of this $p^\mu$ with
the fraction $N^{\rm h}_\gamma$.

\section{Results}\label{sec:results}

All calculations are done for Pb+Pb-collisions at incident beam energy of
$E_{\rm Lab} = 158~A{\rm GeV}$. The sample of collisions has impact
parameters of $b \le 4.5~{\rm fm}$ and only midrapidity-photons ($|y_{\rm
c.m.}| < 0.5$) are included in the analysis. These settings are equivalent
to the WA~98 trigger conditions and detector setups for their ``central''
data set~\cite{Aggarwal:2000ps,Aggarwal:2000th}.

\subsection{Emission times}\label{ssec:times}

Due to formation time effects and the Heisenberg principle, particles cannot
be produced earlier than $t_{\rm prod} \approx E^{-1}$. For photons at
mid-rapidity, where $E = p_\bot$, this gives a lower bound for particle
production times in different $p_\bot$-bins.

Later emission times, though, are well possible. If a particular scattering
occurs preferably at late times, the average emission time of photons may be
shifted to later times.

In this analysis only scatterings between mesons are considered as photon
sources. Mesons have to be produced in a heavy ion collision, so it can be
assumed that meson-meson scatterings occur only after some time. 

\begin{figure} \input{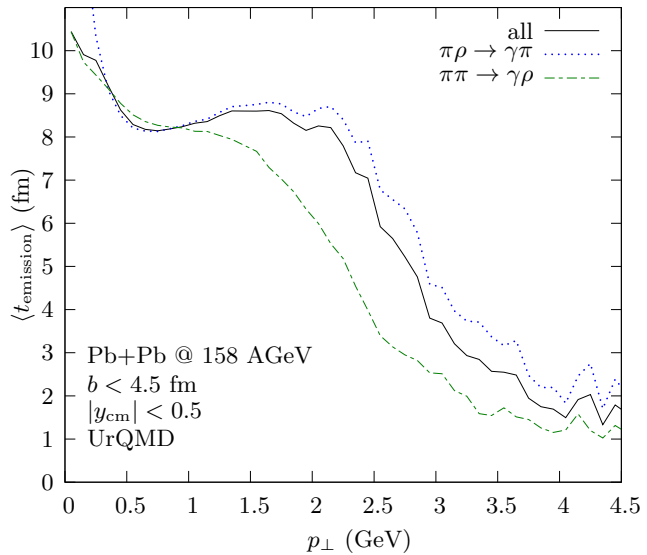} \caption{(Color Online.) Average
emission time of direct photons vs.\ transverse momentum. The curve for
$\pi\pi \rightarrow \gamma\gamma$ is omitted from the plot; the emission
times for that rate are much higher than the range of the plot.}
\label{fig:times} \end{figure}

Figure~\ref{fig:times} shows the average emission times of all photons as
well as of those photons coming from $\pi\rho$- and $\pi\pi$-scatterings.
The complete average is obviously dominated by the processes with $\pi$ and
$\rho$ in the initial state, as is consistent with the results in
section~\ref{ssec:spectra}.

It is eye-catching that the average emission times of the photons deviate a
lot from the hyperbolic behaviour mentioned earlier. Especially in the
intermediate $p_\bot$-region of $1 - 2.5~{\rm GeV}$ it becomes apparent that
photons are produced from decay products that have not been present in the
collision before 5 or 6 fm. $\pi\pi$-scatterings contribute at the same
times, but at lower momenta ($p_\bot \approx 0.5 - 1.5~{\rm GeV}$).

\subsection{Photon spectra}\label{ssec:spectra}

\begin{figure*}
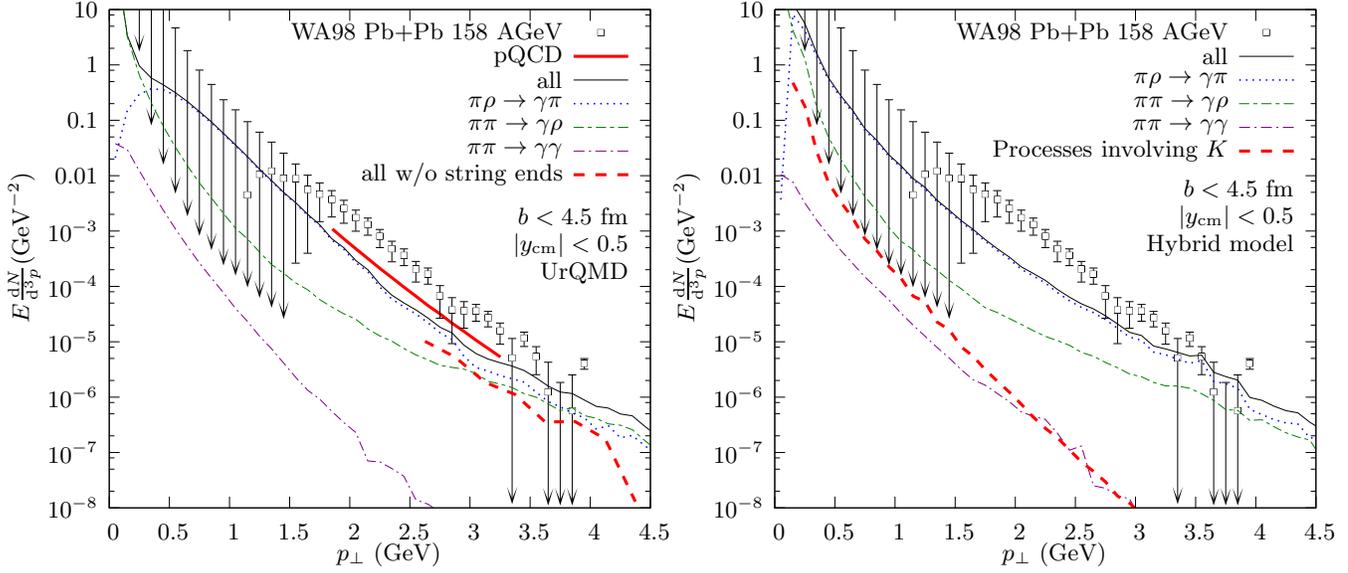
 \input{wa98_cascade_paper} \input{wa98_hydro_paper} \caption{(Color Online.)
Transverse momentum spectra of direct photons from both models. Left: pure
UrQMD, right: Hybrid. The different
charge combinations of the channels have been combined for better clarity.
Experimental data taken from \cite{Aggarwal:2000ps} (``central'' data set,
table IV), arrows indicate data points where the error band extends to
negative yields. In the left figure, a pQCD-spectrum from the initial
proton-proton-collisions \cite{Wong:1998pq,Gale:2001yh} is shown, as well as
the spectrum from pure UrQMD when neglecting the contribution from string
ends (see text).
} \label{fig:cascade}\label{fig:hybrid} \end{figure*}

Figure~\ref{fig:cascade}~(left) shows the direct photon spectra from a pure
transport calculation. The most dominant contribution to the spectra is the
channel $\pi \rho \rightarrow \gamma \pi$, which contributes about on order
of magnitude more than the sum over all other channels between $p_\bot =
0.5~{\rm GeV}$ and $p_\bot = 2.5~{\rm GeV}$.

Emission from all channels is very thermalized at low transverse momenta,
and the $\pi\pi \rightarrow \gamma\rho$- and $\pi\rho \rightarrow
\gamma\pi$- channels contribute significantly at high $p_\bot$, where
pre-equilibrium scatterings play an important role.

As expected, the doubly-electromagnetic channel $\pi\pi \rightarrow
\gamma\gamma$ is very much suppressed over the whole $p_\bot$-range.

The relative contributions of the different channel are very similar in the
case of a hybrid-model calculation. Figure~\ref{fig:hybrid}~(right) shows the
spectra obtained in that case. Here, the dominance of the $\pi\rho$-channel
is as pronounced as in the pure transport calculation. Overall,
contributions from that channel and the $\pi\pi$-channel look rather similar
to the UrQMD-case above a momentum of about $p_\bot \approx 1~{\rm GeV}$,
because in this regime the non-equilibrium phases contribute more than the
high-density hydro phase.

In UrQMD, the leading particles from a string have a reduced cross-section
during their formation time. The effects of neglecting photons coming from
collisions of string ends is shown in Figure~\ref{fig:hybrid}. The
difference becomes important at high $p_\bot$, where photons from initial
hard pQCD-scatterings play an important role.

For comparison, we show the pQCD-spectrum in Figure~\ref{fig:hybrid} (taken from
\cite{Gale:2001yh}) where an intrinsic $\langle k_\bot^2\rangle \approx
0.9~{\rm GeV^2}$ was used. One clearly observes that pQCD photons are a
dominant source of photons at high $p_\bot$, especially if the (slightly
artificial) contribution from string ends is removed from the spectrum.

\begin{figure} \input{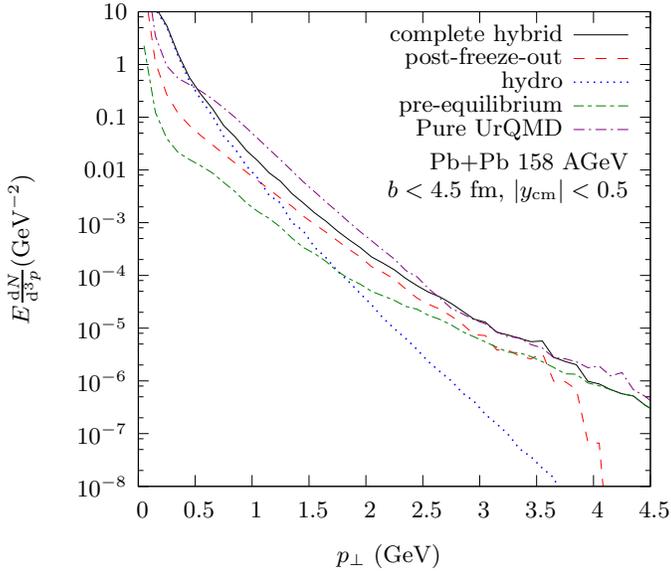} \caption{(Color Online.)
Transverse momentum spectra of direct photons compared between both
approaches.} \label{fig:compare} \end{figure}

For a comparison between hybrid- and transport-cal\-culation we restrict the
analysis to those channels that are present in both models.
Figure~\ref{fig:compare} shows spectra for $\pi\pi \rightarrow \gamma \rho$
plus $\pi\rho \rightarrow \gamma\pi$ for pure transport and the complete
hybrid model. The different stages to the hybrid model (pre-equilibrium,
hydro, post-freeze-out) are seperately plotted.

The yield at low $p_\bot < 1~{\rm GeV}$ is dominated by radiation
from the hydro phase. Above that value, most photons come from the
post-freeze-out and pre-equilibrium stages. It is interesting to see that
the photon yield in the range $1~{\rm GeV} < p_\bot < 2.5~{\rm GeV}$ is
dominated by photons coming from post-freeze-out scatterings. This is
consistent with the observation of late emission times in that
$p_\bot$-range as shown in Figure~\ref{fig:times}. 

\section{Summary}

In this work, we have shown the direct photon spectra for central
Pb+Pb-collisions at top SPS energies. We have used and compared two
different models, both of which give similar results. UrQMD, as a purely
hadronic transport model, underpredicts the data from WA~98 below $p_\bot =
3.5~{\rm GeV}$. The hybrid calculation, where the high-density phase of the
evolution is described by hydrodynamics, also underpredicts the data in that
range.

In both cases, processes with $\pi$ and $\rho$ in the initial state dominate
the intermediate and low momentum regions and processes with two $\pi$
become important at high momenta, where early scatterings dominate.

Emission times for direct photons at intermediate transverse momenta are
found to be very late, and so this $p_\bot$-range is in the hybrid model
dominated by photons from the late post-freeze-out stage.

At very low $p_\bot$, the hydro phase contributes much more to the photon
spectra than the transport phase does. Here, the only significant difference
between both models is seen.

\section{Conclusions}

The hadronic models used in the current work cannot reproduce the direct
photon spectra as measured by WA~98. The reason for this remains unclear; it
could be that the implemention of more channels is necessary. It
is also possible that using an EoS with phase-transition to a QGP will yield
significantly higher spectra. The addition of photons from hard
pQCD-scatterings will contribute significantly, though not sufficiently, to
the final spectrum. Most probably, each of these points will
contribute to the missing photons.

\section{Outlook}

The work presented here lays the foundation for many further investigations.
First, photons from initial hard pQCD-scatterings have to be added. Other
collisions with high $\sqrt{s}$ have to be treated with pQCD instead of
hadronic cross-sections as well. The
channels that are implemented in both stages of the model (transport and
hydrodynamics) have to be aligned, so that a comparison does not have to be
confined to the smallest common denominator. Also, the list of channels in
both cases has to be extended beyond what is implemented so far. Scatterings
involving the $\eta$-meson as well as baryonic processes should be among
those that are added next.

Also, the existing channels can be improved: Until now, the channels that
involve a $\rho$-meson as outgoing particle always assume that the
$\rho$-meson is produced at its pole mass, which clearly does not have to be
the case.

On the part of the hydro evolution, it is of course of great interest to
study the emission of direct photons from thermal partonic processes, i.e.\
the inclusion of an EoS containing a phase transition to a QGP is necessary.
Comparisons to data from RHIC and predictions for the new FAIR-facility will
be accessible then.

\section{Acknowledgements}

This work has been supported by the Frankfurt Center for Scientific
Computing (CSC), the GSI and the BMBF. The authors thank Hannah Petersen for
providing the hybrid- and Dirk Rischke for the hydro-code. B.\ B\"auchle
gratefully acknowledges support from the Deutsche Telekom Stiftung and the
Helmholtz Research School on Quark Matter Studies. This work was partially
supported by the Hessian LOEWE initiative Helmholtz International Center for
FAIR.

\end{document}